\documentclass[lettersize]{emulateapj}

\usepackage{natbib}

\newcommand{\lya}{Ly$\alpha$}
\newcommand{\heii}{\ion{He}{2}}
 
\newcommand{\beq}{\begin{equation}}
\newcommand{\eeq}{\end{equation}}
\newcommand{\bc}{\begin{center}}
\newcommand{\ec}{\end{center}}
\newcommand{\bfig}{\begin{figure}}
\newcommand{\efig}{\end{figure}}
\newcommand{\snr}{\ensuremath{\mathrm{S/N}}}

\newcommand{\D}{\ensuremath{\Delta}}
\newcommand{\tautrue}{\ensuremath{\tau^{\mathrm{true}}}}
\newcommand{\taufit}{\ensuremath{\tau^{\mathrm{est}}}}
\newcommand{\Ctrue}{\ensuremath{C_{\mathrm{true}}}}
\newcommand{\Cfit}{\ensuremath{C_{\mathrm{est}}}}
\newcommand{\gtrue}{\ensuremath{\gamma^{\mathrm{true}}}}

\newcommand{\tdr}{temperature-density relation}
\newcommand{\fmean}{\ensuremath{\langle F \rangle}}

\slugcomment{Second revised draft; Submitted to ApJ}

\begin{document}
\title{Systematic Continuum Errors in the Lyman-Alpha Forest and \\
  The Measured Temperature-Density Relation}
\author{Khee-Gan Lee}
\affil{Department of Astrophysical Sciences, Princeton University, Princeton, New Jersey 08544, USA}
\email{lee@astro.princeton.edu}

\begin{abstract}
Continuum fitting uncertainties are a major source of error in estimates of
 the temperature-density relation (usually parametrized as a power-law, 
$ T  \propto \Delta^{\gamma - 1} $) of the inter-galactic medium (IGM) through the flux
probability distribution function (PDF) of the Lyman-$\alpha$ forest.
Using a simple order-of-magnitude calculation, we show that few percent-level systematic
errors in the placement of the quasar continuum due to e.g.\ a uniform
low-absorption Gunn-Peterson component, could lead to errors in $\gamma$ 
of order unity. This is quantified further using a simple semi-analytic
model of the \lya\ forest flux PDF. We find that under-(over-)estimates in the continuum level
can lead to a lower (higher) measured value of $\gamma$. By fitting models to mock 
data realizations generated with current observational errors, we find that
continuum errors can cause a systematic bias in the estimated temperature-density relation 
of $\langle \delta(\gamma) \rangle \approx -0.1$,
while the error is increased to $\sigma_{\gamma}
\approx 0.2$ compared to $\sigma_{\gamma} \approx 0.1$ in the absence of
continuum errors. 
\end{abstract}

\keywords{intergalactic medium --- quasars: absorption lines ---
methods: data analysis}

\section{Introduction}
Over the past two decades, the Lyman-$\alpha$ (\lya) forest in the line-of-sight
to distant quasars has emerged as one of the 
most important probes of the high-redshift ($ z > 2$) universe
\citep[see, e.g.,][]{croft+98,mcd+00,croft+02,zald+03,mcd+05}. 
This has been underpinned by theoretical advances
that established the \lya\ forest as arising from neutral hydrogen
embedded in a warm 
photo-ionized inter-galactic medium (IGM), tracing the density fluctuations due
to gravitational instability in hierarchical clustering cosmological models
\citep[see, e.g.,][]{cen+94,mirald+96,dave+99,theuns+98} .

In recent years, the \lya\ forest is increasingly being used
to gain a more detailed understanding of the inter-galactic medium.  
In particular, the thermal history of the IGM holds
the key to understanding hydrogen and \heii\ reionization at 
$z > 6$ and $z \sim 3$, respectively.
For an underlying density distribution, $\Delta(x)=\rho(x) / \bar{\rho}$,
 the astrophysics of the IGM mediates
the optical depth of the \lya\ forest. 

In the standard photoionization equilibrium model of the \lya\ forest \citep{gp65}, 
the thermal properties of the IGM 
are usually described in terms of 3 main parameters: the ionization rate of the 
photoionizing background radiation, $\Gamma$, the temperature of the gas at 
mean density, $T_0$, and the temperature-density 
relationship\footnote{The 
temperature-density relationship of the IGM is often called the `equation of
state'. 
This is technically incorrect as the equation of state of the IGM is that of a perfect gas, 
thus in this Letter we do not use this term.}
approximated as a power-law, $T  = T_0 \Delta^{\gamma - 1}$. 

Various authors have placed constraints on the 
background ionization rate, $\Gamma$, of the IGM 
using the effective optical depth, $\tau_\mathrm{eff}$, of 
the \lya\ forest \citep{bolton+05,fg+08b} and the 
quasar proximity effect \citep{scott+00,dall+09}.
Constraints on $T_0$ have been made using detailed 
line-profile fitting of individual \lya\ absorption lines 
from high-resolution spectra \citep[see, e.g.,][]{schaye+00,becker+11}.

Meanwhile, the probability
distribution function (PDF) of the transmitted \lya\ forest flux \citep{jen+ost91} 
has been used to place constraints
 on the \tdr\ of the IGM.
Using this technique, \citet{becker+07}, \citet{bolton+08}, and \citet{viel+09}
have recently found evidence of a highly inverted 
 temperature-density relation, $\gamma \sim 0.5$, in the \lya\ forest at $z \sim 2-3$. 
While some theories \citep[e.g.,][]{furl+oh08} predict a mildly-inverted 
temperature-density relation ($\gamma \approx 0.8$) 
as a consequence of inside-out \heii\ reionization, 
the amount of energy that needs to be injected into the IGM to obtain $\gamma \sim 0.5$
is inconsistent with the observational constraints on heating sources that those redshifts \citep{mcquinn+09}.
 
 Although various systematics such as continuum fitting
 errors, noise and metal line contamination can bias the interpretation of the flux PDF, 
 high-resolution ($R \equiv \lambda / \Delta \lambda \sim 10^4$) and high signal-to-noise 
 ($\snr \gtrsim 50$ per pixel) \lya\ forest spectra can ameliorate these 
effects \citep[see, e.g.,][]{kim+07}. 
 With such data, metal lines in the \lya\ forest region can be directly identified and removed, while 
 the low pixel noise allows precise determinations of the continuum from 
 the peaks of the observed \lya\ transmission, with random errors as low as $1-2 \%$.

However, depending on the underlying properties of the IGM, 
 a uniform Gunn-Peterson absorption component\footnote{While this 
 term may be reminiscent of the obsolete picture
of the IGM as a two-phase medium consisting of 
dense and cool \lya\ clouds embedded in a hot, tenuous inter-cloud medium
\citep{sargent+80}, in this paper we merely use this term to refer generically to 
a uniform or large-scale low-absorption component in the \lya\ forest transmission field.}
 could exist in
 the \lya\ forest even at relatively low redshifts ($z \sim 3$).
 In such a case, a continuum fitted to the transmission peaks of the observed \lya\ forest flux 
 could underestimate the continuum, since few of the peaks would 
 reach the true quasar continuum. 
 This \emph{systematic} continuum bias could exist even in high-\snr\ spectra which have small 
 \emph{random} errors in the pixel fluxes.
 Workers in this field are aware of this possibility:
 \citet{bolton+08} tested this by fitting their flux PDFs with the continuum
 raised by 1.5\% and 5\%, but found that these provided worse fits to their data. 
 \citet{becker+07} and \citet{viel+09} made the 
 continuum level a free parameter in their likelihood analysis, although neither
 found a significant continuum bias.
 
 However, none of the aforementioned studies tried to 
directly estimate the amount of continuum bias at low-redshift ($z \lesssim 3-4$), 
nor is it well-understood
 how such a bias could affect measurements of the IGM \tdr.
 
In a paper measuring the effective optical depth $\tau_{\mathrm{eff}}$ of the \lya\ forest, 
\citet{fg+08} did attempt to constrain the amount of bias in their continuum fits
by hand-fitting mock spectra derived from numerical simulations.
They found that even at $z = 3$, continua fitted to the peaks of the transmission 
underestimate the level of the 
continuum by 5\%, although they assumed an extreme value of $\gamma = 1.6$ in their 
mock spectra.
While this was the temperature-density relation calculated for a relaxed IGM following
hydrogen reionization \citep{hui+gned97}, 
it is not expected to be valid during the epoch of \heii\ reionization at
$z \lesssim 3$ \citep{furl+oh08, mcquinn+09}.
Since $\tau_{\mathrm{eff}} = - \ln( \langle F \rangle)$, a systematic bias in the estimated continuum level
leads to errors in $\tau_{\mathrm{eff}}$ at approximately the same level 
(for example, a 2\% underestimate in the continuum would lead to roughly a 2\% overestimate
in $\tau_{\mathrm{eff}}$).
In contrast, such continuum errors would bias the estimated temperature-density relation 
in a more complicated manner, which is not well-understood. 
This short paper is intended to investigate this bias in a simple, easily-reproducible, 
manner.

In Section~\ref{sec:simple} we first make a simple back-of-the-envelope calculation which
indicates that systematic biases in the continuum fitting of even a few percent
could cause large errors in the estimated \tdr.
Section~\ref{sec:analysis} then introduces a simple semi-analytic model of the 
\lya\ forest flux PDF to quantify this error, followed by a 
discussion (Section~\ref{sec:conclusion}) of these biases 
and some methods to correct for them.

\section{Simple Estimates} \label{sec:simple}
In this section we make an order-of-magnitude 
calculation of the bias
in the IGM temperature-density relation that could potentially 
arise from systematic errors in the placement of the quasar continuum level.
We first define the fractional continuum error,
\begin{equation}\label{fc_def}
f_c = \frac{\Cfit - \Ctrue}{\Ctrue},
\end{equation}
where \Ctrue\ is the true quasar continuum in the \lya\ forest, while \Cfit\ is the 
estimated continuum.

The flux transmission through the IGM $F$ is related to the intervening 
optical depth $\tau$ by $F = \exp(-\tau)$, while $\tau$ is in turn
related to the underlying matter distribution, 
$\Delta(x) \equiv \rho(x) / \bar{\rho}$.

For this calculation we use the fluctuating Gunn-Peterson approximation
\citep[FGPA, see, e.g.,][]{croft+98},
\beq \label{eq:fgpa}
\tau = \tau_0 \D^{2-\alpha},
\eeq
where $\D \equiv \D(x)$, $\alpha = 0.7(\gamma-1)$, 
and $\tau_0$ is a factor which includes astrophysics such as the 
\lya\ recombination coefficient and the photoionizing UV background $\Gamma$.
Here we assume $\tau_0$ to be constant, so that the only unknown parameter 
of the IGM is the exponent $\alpha$ in the equation of state.
 
We rearrange Equation~\ref{eq:fgpa} to
\beq
\alpha = \frac{\ln \tau_0 - \ln \tau}{\ln \D} + 2.
\eeq
The derivative of $\alpha$ with respect to $\tau$ is then
\beq
\frac{d\alpha}{d\tau} = - \frac{1}{\tau \ln \D}.
\eeq
This allows us to estimate the error
 in the derived equation of state,
$\delta \alpha \equiv \alpha^{\mathrm{est}} -\alpha^{\mathrm{true}} $
(the superscripts 
`est' and `true' refer to the estimated
and true underlying quantities respectively), 
arising from a systematic error in the measured optical depth
$\delta \tau $ for a given matter overdensity $\D$:
\begin{eqnarray}\label{eq:dalpha}
\delta \alpha &\sim& -\frac{1}{\ln \D} \frac{\delta \tau}{\tau} \nonumber \\
&\sim& \frac{\delta \tau}{\tau}.
\end{eqnarray}
In the second line, we have approximated  $-\ln \D $ to be of order unity,
since the low-absorption regions of the IGM are in the underdense
($\Delta < 1$) parts of the universe. 

Now consider the effects of a misunderestimated quasar continuum level
\Cfit, where \Ctrue\ is the true continuum level. 
\citet{fg+08} have shown that the true optical depth \tautrue\ is related to the measured
optical depth corresponding to the underestimated continuum, \taufit\ by 
\begin{eqnarray} \label{eq:dtau}
\delta \tau &=& \taufit - \tautrue \nonumber \\
&=& \ln\left(1 + \frac{\Cfit - \Ctrue }{\Ctrue}\right) \nonumber \\
&\simeq& f_c,
\end{eqnarray}
where in the last line we have substituted in Equation~\ref{fc_def} and assumed that it
is small ($|f_c| \lesssim 0.1$).

Inserting Equation~\ref{eq:dtau} into Equation~\ref{eq:dalpha}, and
using the fact that in the low-absorption regions 
$F = \exp(-\tau) \simeq 1 - \tau$, we get
\beq
\delta \alpha \sim \frac{f_c}{1 - F}.
\eeq

As the strongest constraints on the \tdr\  of the IGM come from the 
low-absorption/high-transmission ($1 - F \sim \%$) pixels of the 
\lya\ forest, systematic errors in the continuum estimate of just
a few percent can cause large ($\delta \gamma \sim \delta \alpha \sim 1$)
errors in the estimated value of $\gamma$.

\section{Mock Analysis}\label{sec:analysis}
In this section, we generate theoretical flux PDFs of the \lya\ forest
using a simple semi-analytic model, and quantitatively study the errors introduced into 
the estimated \tdr\ caused by systematic
errors in the \lya\ forest continuum. 

\subsection{Toy Flux PDFs} 

\begin{figure}
\epsscale{1}
\plotone{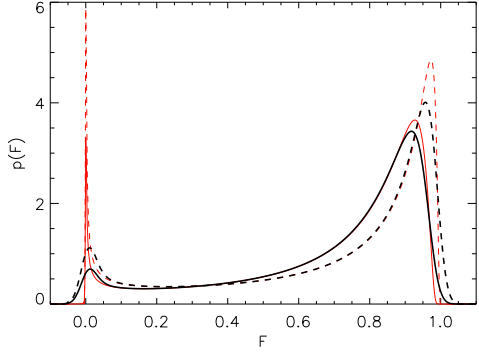}
\caption{\label{fig:rawpdfs}
Unbinned flux PDFs generated from Eq.~\ref{eq:fpdf}, with 
$\gamma=0.5$ (thin red dashed line) and $\gamma=1.6$ 
(thin red solid line). The flux PDFs are normalized
to $\langle F \rangle = 0.70$.
 The thick black lines denote
the PDFs after being smoothed by a Gaussian kernel corresponding
to $\snr=50$ in our simplified noise model.
}
\end{figure}

\begin{figure}
\epsscale{1.}
\plotone{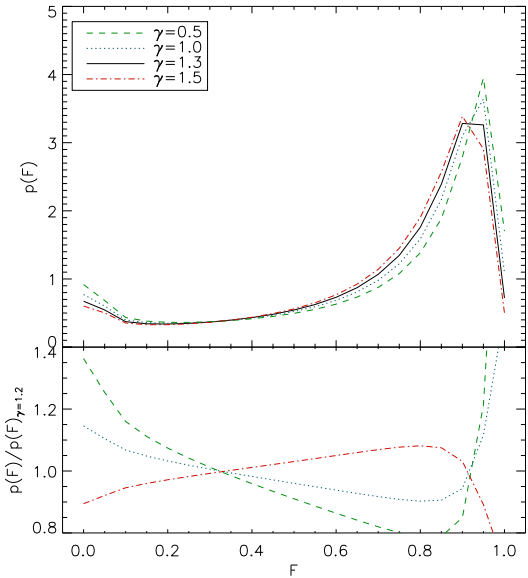}
\caption{\label{fig:binnedpdfs}
(Upper panel) Binned flux PDFs generated from Eq.~\ref{eq:fpdf} with 
different temperature-density
relations, $\gamma=0.5$ (green dashed line), $\gamma=1.0$ (blue dotted line),
$\gamma=1.2$ (black solid line), and $\gamma=1.6$ (red dot-dashed line). 
The lower panel shows the ratio of the various flux PDFs
with respect to the $\gamma = 1.0$ model. The relative differences 
are similar to those presented in \citet{bolton+08}.
}
\end{figure}

We first generate theoretical flux PDFs by using as a starting point the semi-analytic density
PDF from \citet{mhr00} (hereafter MHR00):
\beq \label{eq:pmhr}
p(\Delta) d\Delta = A \exp{\left[\frac{-(\Delta^{-2/3} - C_0)^2}{2 (2\delta_0 /3)^2}
\right]} \Delta^{-\beta} d\Delta,
\eeq
where $\Delta$ is the matter density, while $A$, $\beta$, $C_0$, and 
$\delta_0$ are redshift-dependent parameters interpolated from the values 
published in MHR00. 
At $z=3$, the corresponding values are $A = 0.558$, $\beta=2.35$,
$C_0 = 0.599$, and $\delta_0 = 1.90$.

The corresponding \lya\ forest flux PDF is then derived by substituting in $\Delta(\tau)$ using the
 FGPA (Equation~\ref{eq:fgpa}) and 
then the relation $F = \exp{(-\tau)}$:
\begin{eqnarray} \label{eq:fpdf}
p(F) dF &=& \frac{dF}{F} \frac{A}{\tau_0 (2 - \alpha)}
\left( \frac{-\ln{F}}{\tau_0} \right)^{\frac{\alpha-1-\beta}{2 - \alpha}}
\nonumber \\
& & \times \exp{\left\{ \frac{-\left[(-\ln{F}/\tau_0)^{-\frac{2}{3(2-\alpha)}}
	- C_0 \right]^2}{2 (2\delta_0/3)^2}  \right\}}.
\end{eqnarray}
This is then normalized such that $\int_0^1 p(F) dF = 1$.

The parameters in Equation~\ref{eq:fpdf} 
that characterize the IGM are $\alpha$  and $\tau_0$. 
$\tau_0$ is a function that depends on the background photoionization rate
in the IGM and the temperature at mean density, but for the purposes
of this paper we adjust it to yield a fixed value of mean optical 
depth $\langle F \rangle (z) = \exp{(-\tau_{\mathrm{eff}}(z))}$ 
for a given redshift. 
Thus, at fixed redshift the only free parameter in this model is 
 $\gamma = 1 + \alpha/0.7$ which parametrizes
  the temperature density equation-of-state.

Note that this model does not correctly account for thermal broadening 
of the \lya\ forest lines, nor peculiar velocities. 
However, \citet{bolton+08} have shown from hydrodynamical simulations 
that the flux PDF is not sensitive to $T_0$, which affects the
\lya\ forest primarily through thermal broadening and by
changing the Jean's smoothing scale. 
Peculiar velocities need to be included to obtain the correct flux PDF shape, 
but they arise primarily from gravitational collapse of large-scale
structure. Thus, we do not expect the omission of peculiar velocities
to seriously affect the relative behavior of the flux PDF with changes to the 
temperature-density relation.

We show the flux PDFs calculated for 
2 extreme values, $\gamma=0.5$ and $\gamma=1.6$
in Figure~\ref{fig:rawpdfs}. 
$\gamma=0.5$ corresponds to a highly inverted temperature-density
relation, which had been detected by \citep{becker+07, bolton+08}.
While $\gamma=1.6$ is a value calculated by \citet{hui+gned97} for a
post-reionization relaxed IGM, which should be valid in the epoch between 
the end of hydrogen reionization and prior to \heii\ reionization, 
$ 3 \lesssim z \lesssim 6$.  

Figure~\ref{fig:rawpdfs} illustrates the effects of changing $\gamma$:
at fixed temperature at mean density $T_0$,
 lowering $\gamma$ increases the temperature in the underdense
($\D < 1$) regions of the IGM. 
This decreases the hydrogen recombination rate, therefore
reducing the \ion{H}{1} optical depth and increasing the transmission
in those regions. The differences are apparent in
the  $F \gtrsim 0.5$ regions of the PDF, where the peak is shifted
to higher $F$. 
The probability of having pixels with close to 
100\% transmission decreases as $\gamma$ goes up;
at $\gamma=1.6$, the probability drops to zero by $F \simeq 0.98$ 
(in our model) thus
leading to Gunn-Peterson absorption at the 2\% level. 
In this case, the lack of transmission peaks reaching the true 
continuum level is likely to lead to an underestimate 
of the continuum, potentially biasing the estimate of $\gamma$.
Note that the moderate-absorption regions of 
the PDF ($0.2 \lesssim F \lesssim 0.6$)
do not vary much with $\gamma$. This agrees with \citet{bolton+08}, 
who found that using only these moderate-absorption pixels 
significantly weakens constraints on $\gamma$.

The effects of finite signal-to-noise are introduced by smoothing 
the flux PDF from Equation~\ref{eq:fpdf} with a Gaussian kernel.
 We use a constant smoothing length of
$\sigma_{sm} = 1/50$ to simulate an average $\snr= 50$ per 
pixel, a typical value for high-resolution and high-$\snr$ \lya\ forest
spectra.
This is an acceptable simplification, as Table~3 in \citet{mcd+00}
shows that the typical pixel noise in real data is 
roughly constant across the different flux bins.
Note that the presence of noise scatters some pixels to 
$F \gtrsim 1$, changing the shape of the flux PDFs
(black curves in Figure~\ref{fig:rawpdfs}).

We then bin the flux PDFs in the fashion of \citet{mcd+00} and 
\citet{kim+07}: the `noisy' flux PDF is divided into 21
bins with size $\Delta F = 0.05$ in the range $0 < F < 1$.
The `noisy' portions of the flux PDF
with $F < 0$ and $F > 1$ are transferred to the $F=0$ and $F = 1$ bins, 
respectively. 

In Figure~\ref{fig:binnedpdfs}, we show the binned PDFs for different 
temperature-density relations. The overall trend of the unbinned PDFs
are similar to that in the unbinned case: 
the high-transmission peak of the PDF 
shifts to larger $F$ as $\gamma$ is decreased. In addition, the 
number of pixels in the $F=1.0$ bin increases with decreasing $\gamma$.
The overall shape of the flux PDFs are in broad agreement with other
theoretical flux PDFs published in the literature, including 
\citet{mcd+00}, \citet{bolton+08}, and \citet{white+10}. 
However, note that we are unable to reproduce the exact shapes of the 
flux PDFs as seen in the \citet{kim+07} data, 
nor from hydrodynamical simulations \citep[e.g.,][]{bolton+08}:
our PDFs tend to peak at lower values of $F$ for a given value of $\gamma$, 
while the $\gamma=0.5$ 
PDF here does not have the same shape in the high-transmission end obtained 
from the \citet{bolton+08} hydrodynamical simulations, 
in which the PDF peaks in the $F=1.0$ bin.
The MHR-FGPA model also appears to underpredict the number of
low-transmission pixels ($F\approx 0$) in comparison with the 
\citet{bolton+08} simulations.

In the lower panel of Figure~\ref{fig:binnedpdfs}, we plot the ratio of 
the flux PDFs with respect to that with $\gamma = 1.2$. 
These appear 
similar to the analogous plots shown in the lower panels of Figures~2 
and 4 in \citet{bolton+08}. 
The main differences are that the flux PDFs in the hydrodynamical
simulations pivot at $F\approx 0.1$, while those in our model 
pivot at a higher value of $F\approx 0.3$. 
Nevertheless, the \emph{relative} behavior of our model PDFs with
respect to $\gamma$ appear similar to those in the \citet{bolton+08}
 hydrodynamical simulations. 
Since the primary effect of systematic continuum errors is to 
rescale the flux PDF along the abscissa, 
the fact that our toy model correctly reproduces the relative 
changes in the PDF with respect to $\gamma$ means that it can be used
to study continuum errors, even if it should not be used to make direct
comparisons with data.

\subsection{Effect of Continuum Errors}

\begin{figure}
\epsscale{1.}
\plotone{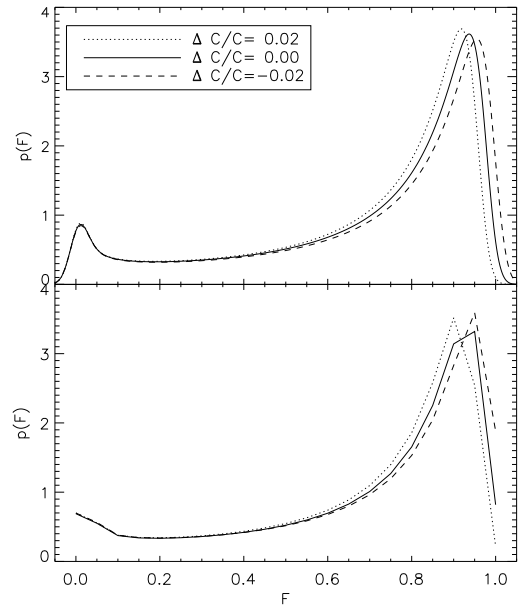}
\caption{\label{fig:conterr}
The effect of systematic continuum errors on the \lya\
forest flux PDF.
The top panel shows the unbinned flux PDFs generated with 
$\gamma = 1.2$ and $\snr = 50$, with no continuum error (black line), 
2\% overestimated continuum (dotted line), and 2\% underestimated
continuum(dashed line). The lower panel shows the corresponding
flux PDFs in $\Delta F=0.05$ bins. 2\% errors in the continuum
determination can drastically change the shape of the binned flux PDF.
}
\end{figure}  

\begin{figure}
\epsscale{1.}
\plotone{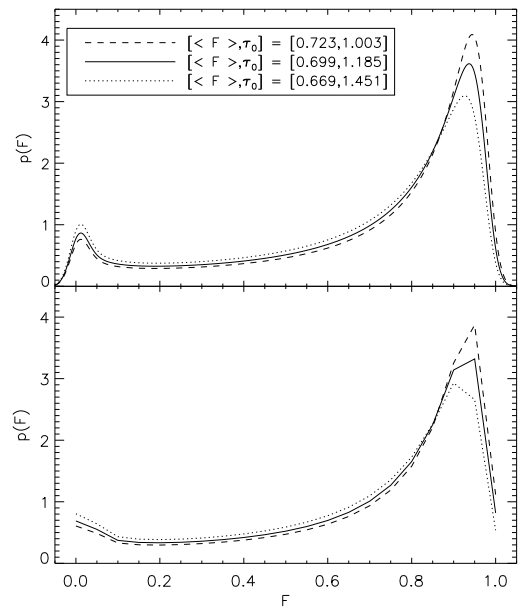}
\caption{\label{fig:meanflux}
The effect of changing $\tau_0$ (or equivalently $\fmean$)
on the flux PDF. The top panel shows the unbinned flux PDFs generated with 
$\gamma = 1.2$ and $\snr = 50$, with $\tau_0$ adjusted to match the effective 
mean-flux of the PDFs with the corresponding line-styles in Figure~\ref{fig:conterr}. 
The lower panel shows the corresponding
flux PDFs in $\Delta F=0.05$ bins. $\tau_0$ changes the flux PDF
differently from the temperature-density relation or continuum errors. No
continuum errors have been applied in these PDFs.
}
\end{figure}  

In this section, we study the effect of systematic continuum biases
on the value of $\gamma$ measured from the mock flux PDFs
described above,
and the corresponding errors on this estimate.
This mock analysis is carried out at a fixed redshift of $z=3$, 
an epoch at which Gunn-Peterson absorption is usually assumed to 
be negligible, 
but could account for as much as 4-5\% of the observed flux
\citep{gial+92,fg+08}.
All mock PDFs are set to a mean flux of 
$\langle F \rangle(z=3)=0.70$ \citep{meik+white04}, and the model parameters
from Equation~\ref{eq:pmhr} are set to $z=3$.

Systematic continuum errors are introduced into the flux PDFs 
by multiplying the flux scales with 
the factor $1 + f_c$ prior to binning --- positive values of $f_c$ denote
an overestimate with respect to the true continuum, and vice versa.
The dotted and dashed curves in Figure~\ref{fig:conterr} illustrate
the effects of these errors on a flux PDF with $\gamma=1.2$. 
Clearly, just $2\%$ systematic errors in the continuum estimation can 
dramatically change the shape of the binned flux PDFs:
with a 2\% underestimate of the continuum, the number of unabsorbed 
pixels ($F=1.0$) increases by a factor of two.
Note that the continuum errors also change the measured mean-flux
$\langle F \rangle$. For the cases with $f_c = [-0.02, 0, 0.02]$, the
effective mean-fluxes are $\fmean = [0.723, 0.699, 0.669]$, respectively. 
The scatter in these mean-flux values are at the same
level as the current observational uncertainty of several percent 
at $z \approx 3$. 

 To investigate if the mean-flux might be degenerate
with $\gamma$ and $f_c$, in Figure~\ref{fig:meanflux} we show $\gamma=1.2$ flux PDFs with
$\tau_0$ adjusted to give $\fmean$ values corresponding to the curves in 
Figure~\ref{fig:conterr}, without continuum errors included.   
The corresponding values used to generate these PDFs 
are $\tau_0 = [1.003, 1.185, 1.451]$ for $\fmean = [0.723, 0.699, 0.669]$,
respectively. 
We see that a smaller $\tau_0$ or higher $\fmean$ can shift the high-transmission peak 
of the flux PDF to larger $F$, analogous to lowering $\gamma$.
However,
the relative ratio of high-absorption/low-transmission pixels ($F < 0.85$), 
with respect to the low-absorption/high-transmission end ($F > 0.85$)
behaves differently with changes in $\tau_0$ compared to when 
$\gamma$ or $f_c$ is varied (c.f. Fig.~\ref{fig:binnedpdfs} 
and Fig.~\ref{fig:conterr}). 
This suggests that the degeneracy of $\tau_0$ with respect to 
$\gamma$ and $f_c$ can be broken in an analysis of the full flux PDF,
but nevertheless $\tau_0$ needs to be taken into account.

To quantify the effects on the estimated temperature-density relation, 
the flux PDFs with different combinations of $[\gamma, f_c]$ are
then compared with mock data PDFs with similar noise properties to 
the \citet{kim+07} flux PDFs, which is the best existing data set.
However, the bins of the flux PDF are highly correlated, 
thus one needs a covariance
matrix in order to generate random samples with the correct covariances. 
As our semi-analytic model does not actually result in simulated 
\lya\ forest sightlines, it is problematic to estimate the necessary covariances.
Instead, we use the same covariance matrices described in 
\S4.1 of \citet{bolton+08}:
 the covariance matrix (kindly provided by James Bolton) derived from 
their $z=2.94,\gamma=0.44$ simulations are used to estimate 
the correlation coefficients between different flux bins. 
This is then used in conjunction with the errors 
(i.e. diagonal covariance terms) in the \citet{kim+07} data to estimate the 
cross-terms of the final covariance matrix.
We then carry out Cholesky decomposition on this covariance matrix to generate
random correlated errors, which are then applied to the model flux PDFs.
This process assumes that all the model covariances are the same as 
those in \citet{bolton+08},
but this should be reasonable considering the 
approximate nature of our analysis.

We can then quantify the effects of continuum bias by sampling
the reduced Chi-squared between each mock data PDF, and different
model curves
with various combinations of $\gamma$ and $f_c$. Since the bins are
highly correlated, the expression for the Chi-squared is:
\beq
\chi^2 = 
[d_i - p_i^{\mathrm{model}}]^T C_{ij}^{-1} [d_i - p_i^{\mathrm{model}}]
\eeq
where $d_i$ are the bins in the mock data PDFs, 
$p_i^{\mathrm{model}}$ is the model flux PDF with some combination of
$\gamma$, $f_c$ and $\tau_0$, while
$C_{ij}$ is the covariance matrix discussed above.
The confidence level of the various parameter combinations can then be 
estimated from $\Delta \chi^2 \equiv \chi^2 - \chi^2_{min}$. 

In our analysis, we regard the \lya\ forest mean-flux, $\fmean$,
(or equivalently within our model, $\tau_0$)
as a nuisance parameter to be marginalized over. 
At each point in the $[\gamma,f_c]$ model parameter space, we
marginalize over the mean-flux in the likelihood, 
$\mathcal{L} \equiv [ (2\pi)^N \det{C_{ij}} ]^{-1} \exp\{-(1/2) \chi^2 \}$.
If we assume the error in the mean-flux is Gaussian, we can then 
marginalize over the possible mean-flux values $\fmean'$:
\begin{eqnarray}
\mathcal{L}(\gamma,f_c|\fmean) &\propto&
 \int^{\infty}_{\infty}
\mathcal{L}(\gamma, f_c, \fmean' ) \nonumber \\
& & \exp \left(- \frac{(\fmean' - \fmean)^2} {\sigma_{\fmean}^2}  \right) 
d\fmean',
\end{eqnarray}
where we have used $\fmean = 0.70$ and $\sigma_{\fmean} = 0.02$ for $z=3$.
Note that we have ignored normalization factors that will be canceled out when we evaluate
$\Delta \chi^2$, and have evaluated the integral using
7-point Gauss-Hermite quadrature.

\begin{figure} 
\epsscale{1.}
\plotone{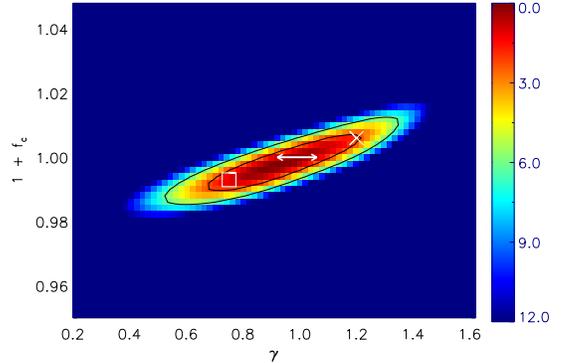}
\caption{\label{fig:chisq}
Likelihood plots for different combinations of $\gamma$ and $f_c$,
 computed with respect to a mock flux PDFs  generated 
with $\gamma=1.0$. 
The contours enclose the 68\% and 95\% confidence intervals,
while the color scale denotes $\Delta \chi^2 \equiv \chi^2 - \chi^2_{min}$.
The square and cross denote the extreme values of $\gamma$ 
plotted in Figure~\ref{fig:binpdf} which
constitute acceptable $\sim 1\sigma$ fits to the $\gtrue=1.0$ mock flux PDF 
(also plotted in Figure~\ref{fig:binpdf}) once continuum errors are considered.
The white horizontal arrow denotes the 68 \% confidence interval for 
$\gamma$ when the continuum is perfectly known, i.e. $f_c = 0.$.
}
\end{figure}

\begin{figure}
\epsscale{1.}
\plotone{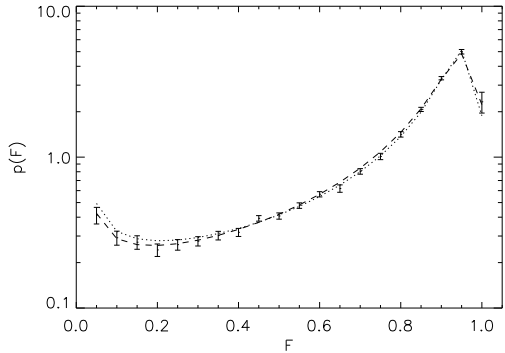}
\caption{\label{fig:binpdf}
Flux PDFs corresponding to Figure~\ref{fig:chisq}.
The error bars denote the mock $\gamma=1.0$ flux PDF 
with simulated scatter, 
against which the likelihoods in Figure~\ref{fig:chisq} were
calculated. 
The dashed- and dotted lines denote model PDFs with 
$[\gamma=0.75,f_c=-0.007] $ and $[\gamma=1.2,f_c=0.006]$, which constitute
acceptable ($\Delta \chi^2 \approx 2$, 68\% confidence level) 
fits to the mock $\gtrue=1.0$ data points.
Their respective positions in $[\gamma,f_c]$ space are shown as 
the square and cross in Fig~\ref{fig:chisq}. 
Note that the ordinate axis is plotted logarithmically in this figure.}
\end{figure}  

Figure~\ref{fig:chisq} shows the likelihood plot for
fits to a mock data PDF generated with $\gtrue=1.0$,
computed with respect to different
 combinations of $\gamma$ and $f_c$.
The contours indicates the 68\% and 95\% confidence intervals 
($\Delta \chi^2 = 2.30$ and $\Delta \chi^2 = 6.17$, respectively).
The error ellipse shows a significant degeneracy 
between $\gamma$ and $f_c$.
For reference, we also show the 68\% confidence level ($\Delta \chi^2 = 1$)
in the case where the continuum is not allowed to vary (horizontal
white arrows).
The range of $\gamma$ which falls within the 68\% confidence 
interval increases significantly to $\sigma_{\gamma} = 0.2-0.3$ once the continuum level is allowed to vary.
Without continuum errors, the 
error\footnote{We use the terms `$1\sigma$ errors' and `68\% confidence intervals'
interchangeably but the latter is the quantity we have really measured.
These two terms would be identical in the case of Gaussian errors.} in the estimated $\gamma$ is $\sigma_{\gamma} \approx 0.05$.

The square and cross symbols in Figure~\ref{fig:chisq} 
denote two extreme points
in the $[\gamma, f_c]$ parameter space
 that fall within the
68\% confidence interval of the underlying 
flux PDF with $\gtrue = 1.0$.
The binned flux PDFs for these two combinations of $[\gamma, f_c]$
 are juxtaposed
with the underlying PDF in Figure~\ref{fig:binpdf}.
The similarity of these PDFs with significantly different 
temperature-density relations ($\gamma = 0.75-1.2$) illustrates
the effect of continuum errors on the measurement.

\begin{figure}
\epsscale{1.}
\plotone{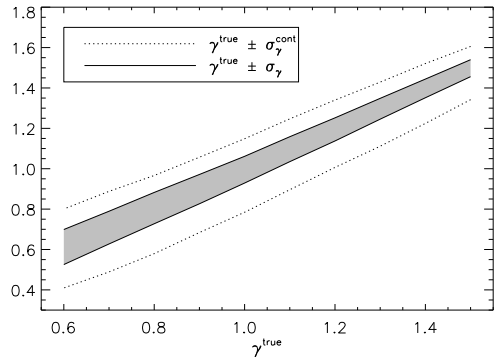}
\caption{\label{fig:eos_errs}
The $1\sigma$ upper- and lower-limits in the value of $\gamma$ estimated from 
mock flux PDFs generated with some underlying value of \gtrue, when continuum
biases are taken into account (dotted lines) and if the continuum is perfectly known (shaded area
enclosed by solid lines). Continuum uncertainties roughly double the error on $\gamma$.
These values are averaged over $40$ mock data realizations for each value of $\gtrue$.
}
\end{figure}

Figure~\ref{fig:eos_errs} shows the average 
estimated value of $\gamma$ as a function of the `true' underlying
temperature-density power-law $\gtrue$ and its associated errors, 
averaged over $40$ Gaussian
 realizations for each value of $\gtrue$.  
The dotted lines denote the average 68\% confidence intervals if systematic continuum errors 
are considered, while the shaded area show the $1\sigma$ 
intervals if 
 the continuum is known perfectly. 
 We see that continuum errors cause
  a significant increase in the error on $\gamma$: 
 at higher temperature-density slopes ($\gamma \sim 1.5$) the errors increase 
 from $\sigma_{\gamma} \approx 0.05$ to $\sigma_{\gamma}^\mathrm{cont} \approx 0.15$;
 at flatter or inverted slopes, the errors are generally larger but still increase 
 from $\sigma_{\gamma} \approx 0.1$ to $\sigma_{\gamma}^\mathrm{cont} \approx 0.2$.
 In general, the errors increase by a factor of $\sim 2$ once we consider continuum 
 errors.
 
\section{Discussion \& Conclusion} \label{sec:conclusion}

\begin{figure}[t]
\epsscale{1.}
\plotone{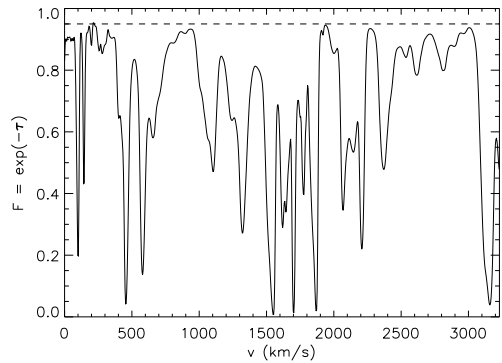}
\caption{\label{fig:cen_spec}
A \lya\ forest sightline from the \citet{cen+chisari10} hydrodynamical simulations extracted at $z = 3.0$
and smoothed to a resolution of $6.7\, \mathrm{km\, s^{-1}}$.
The dashed horizontal line indicates the $F = 0.95$ flux level. 
Even though no pixel noise is present, an attempt to fit the continuum to this mock spectrum
will likely result in a continuum bias of $f_c \approx -0.05$. Courtesy of R. Cen.
}
\end{figure}

In this paper we have quantified the effect
of systematic continuum errors on the temperature-density relation, 
$\gamma$, estimated from the \lya\ forest flux PDF, using a simple toy model
which correctly reproduces the relative changes in the flux PDF with 
respect to $\gamma$.

We found that small systematic errors of just $\sim 1-2\%$ in the overall continuum level 
can bias the estimated $\gamma$ 
to smaller values.
In the absence of continuum errors, the $1\sigma$ errors on the estimated value of $\gamma$
is $\sigma_{\gamma} \approx 0.1$, but with continuum errors this interval is increased
 to $\sigma_{\gamma} \approx 0.2$.

While we have used a simple semi-analytic model 
 to calculate the flux PDFs
(although note that this is the same model used by \citet{becker+07} to 
analyze their spectra), the relative scaling of the resultant flux 
PDF with $\gamma$ is reasonably accurate and thus
the biases discussed here are qualitatively valid. We have also marginalized over 
the mean-flux in our analysis.

In addition, in the simple mock analysis presented here, 
both the data realizations and `theory' flux PDFs are
derived from exactly the same model.
In a real data analysis, uncertainties in the underlying
physical model (e.g.\ gas temperature, UV ionizing background, $\sigma_8$, 
Jeans' smoothing scale etc.), 
and other observational uncertainties such as metal contamination 
must increase the error in $\gamma$ beyond those presented here.

Another point that should be emphasized is that prior to 
the biases in $\gamma$ considered in this paper,
systematic continuum errors are not a symmetric effect. Overestimates
of the continuum ($f_c > 1$) are less likely with high-quality data 
--- an observer is unlikely to place the continuum level 
 significantly above the observed transmission peaks, 
whereas low-level Gunn-Peterson absorption 
can degrade the transmission peaks and lead to 
underestimates of the continuum ($f_c < 0$).
There is thus an additional bias towards smaller values of $\gamma$ 
 due the higher probability of 
underestimating the continuum rather than overestimating it.
In the MHR00 model considered here, the maximum amount of possible 
Gunn-Peterson absorption is in fact fairly limited at $z = 3$: about 2\% 
with $\gamma = 1.6$ (Figure~\ref{fig:rawpdfs}).
Other models could provide even more Gunn-Peterson absorption 
at these redshifts.
An example is shown in Figure~\ref{fig:cen_spec}, which plots a $z=3$
\lya\ forest spectrum extracted from the detailed hydrodynamical simulations
described in \citet{cen+chisari10}. If this were actual data, an observer
would probably underestimate the spectrum by 5\% even 
in the absence of noise.

Recent studies of the flux PDF that claimed a highly inverted \tdr\ 
\citep{becker+07,bolton+08,viel+09} have
considered the possibility of a systematic continuum bias. 
However, \citet{becker+07} used the same semi-analytic MHR00 model for the flux PDF
used in this paper, which may not accurately capture the details of 
the \lya\ forest flux field at a sufficiently level for data analysis. 
\citet{bolton+08} and \citet{viel+09} both analyzed the same data set from \citet{kim+07}:
\citet{bolton+08} checked for continuum errors by comparing their 
calculating likelihoods after rescaling their continua by 1.5\% and 5\%, while
\citet{viel+09} marginalized the continuum in their analysis ---
both concluded that continuum errors were not significant and that
 the inverted \tdr\ was favored.
However, it is also interesting to note that 
\citet{viel+09} arrived at a best-fit continuum error suggesting an \emph{over}estimated 
continuum of 1\% in the \citet{kim+07} data 
(c.f.\ the random errors in the continuum fitting, $\sigma_{\mathrm{fit}} = 1-2 \%$). 
In other words, they favor a continuum which is \emph{lower} than that 
which was actually fitted to the \citet{kim+07} spectra.
In the context of our analysis, this would be consistent with an underestimation of 
$\gamma$.
We also note that \citet{viel+09} had fitted for a redshift-independent continuum error
when analyzing the \citet{kim+07} flux PDFs. 
Since this data set is dominated by lower-redshift ($z< 2.5$)
spectra (which should be less affected by low-level Gunn-Peterson absorption), 
it is possible that the true continuum error cannot be fully accounted for, 
using a redshift-independent approach.
  
Considering the controversial nature of the claims of a highly-inverted
IGM \tdr, we feel a more direct approach towards dealing with continuum bias 
is required that had been done previously.
There are some possibilities: 
\citet{gial+92} had extrapolated a power-law continuum from redwards of the 
quasar \lya\ emission line to estimate the amount of uniform GP
absorption in the peaks of the \lya\ forest. 
Existing data sets will allow much stronger constraints to be placed using
 this method, although it requires the assumption that the mean quasar continuum
slope does not change with redshift. 

Alternatively, when comparing simulations with data, the simulated
sightlines need to be processed through the same continuum-fitting method
as the observed data, i.e. through `forward-modeling'.
\citet{fg+08} estimated continuum biases by
 fitting simulated mock spectra, and comparing these fits
with the underlying continuum. However, they were attempting to 
measure the optical evolution of the \lya\ forest, and used 
 simulations with a fixed value of $\gamma=1.6$.
For this method to account for continuum biases in 
studies exploring a large parameter space, one would need to hand-fit
large sets of mock spectra (ideally including realistic quasar continua) 
covering the explored parameter space. 
This would be time-consuming, but in principle would be a robust
method to account for systematic errors in the continuum fits,
especially if automated continuum-fitting methods are used \citep{dall+09}.

Another avenue for improvement to use larger \lya\ forest data sets,
and hence reduce the errors in the flux PDF and other statistics.
The \citet{kim+07} sample (18 quasars) represent a
significant increase in data size in comparison with \citet{mcd+00} (8 quasars),
but considerably more high-resolution, high-\snr\, spectra than these currently exist.
Increasing the sample size would clearly reduce the errors in the measured flux PDF, 
which would limit the scope of continuum errors to bias the measured
temperature-density relation.

At time of writing, the measurement of the inverted temperature-density relation
has been carried out by the same group of authors \citep{bolton+08,viel+09}
analyzing the same flux PDF data \citep{kim+07}. 
Alternative and independent analyses are urgently required in order to verify this phenomenon, 
which would have important implications on IGM science as well as energetic 
sources in the high-redshift universe.

\acknowledgements{
The author thanks David Spergel, Renyue Cen and Xavier Prochaska for useful
discussions and comments. He is also grateful to James Bolton for
providing covariance matrices from his simulations, and to the anonymous
referee for useful criticisms which have improved the paper. 
}

\bibliography{ms,apj-jour}

\end{document}